\documentclass[
 aip,
 amsmath,amssymb,
 reprint,
]{revtex4-1}
\usepackage{graphicx}
\usepackage{dcolumn}
\usepackage{bm}
\usepackage{multirow}
\usepackage[utf8]{inputenc}
\usepackage[T1]{fontenc}
\usepackage{mathptmx}
\usepackage{color}
\usepackage[page]{appendix}

\usepackage{caption}
\usepackage{subcaption}

\newcommand{\iket}[1]{$\left | {#1} \right\rangle$}
\newcommand{\ket}[1]{\left | {#1} \right\rangle}

\newcommand{\bra}[1]{\left \langle {#1} \right |}
\newcommand{\braket}[2]{\left \langle {#1} | {#2} \right \rangle}

\draft 

\begin{document}

\title{Improving Variational Monte Carlo Optimization by Avoiding Statistically Difficult Parameters}



\author{Scott M. Garner}
\affiliation{Department of Chemistry, University of California, Berkeley, California 94720, USA }
\affiliation{Chemical Sciences Division, Lawrence Berkeley National Laboratory, Berkeley, CA, 94720, USA}
\author{Eric Neuscamman}
\email{eneuscamman@berkeley.edu.}
\affiliation{Department of Chemistry, University of California, Berkeley, California 94720, USA }
\affiliation{Chemical Sciences Division, Lawrence Berkeley National Laboratory, Berkeley, CA, 94720, USA}

\date{\today}

\begin{abstract}
Modern quantum Monte Carlo (QMC) methods often capture electron correlation through both explicitly correlating Jastrow factors and small to mid-sized configuration interaction (CI) expansions.
Here, we study the additional optimization difficulty created by including increasing numbers of CI parameters.
We find evidence that the quality of Variational Mone Carlo (VMC) optimization can be limited by the ability to statistically resolve the CI parameters in the presence of a Jastrow factor.
Although using larger statistical samples can mitigate this issue and bring
an optimization closer to its true minimum, this approach is
computationally intensive.
We present evidence that similar gains to optimization quality can
be had without increasing the sample size by avoiding CI parameters
that are statistically the most difficult to resolve.
Our findings suggest that, in addition to the value of using traditional
selected configuration interaction (sCI) methods to prepare VMC wave
functions, sCI-like methods can play an important role within VMC
by improving the effectiveness of stochastic energy minimization.

\end{abstract}

\maketitle 


\section{Introduction}
\label{sec:intro}

QMC methods offer an alternate approach to traditional quamtum chemistry for finding approximate solutions to the Schr\"{o}dinger equation based on Monte Carlo stochastic sampling of a trial wavefunction.
As has been reviewed many times\cite{luchow2011quantum_Luchow_Review,lester2009quantum_QMC_Review_Lester,carlson2015quantum_QMCReview_Nuclear_Physics,foulkes2001quantum_ReviewQMC_Solids}, the two most common QMC methods are VMC and Diffusion Monte Carlo (DMC), the former of which is the primary focus of this work.
While QMC comes with a non-negligible prefactor cost, the sampling is embarrassingly parallelizable and the asymptotic scaling with system size can
be competitive with modern high accuracy wavefunction methods.
Further, the numeric sampling allows for the inclusion a wide variety of sophisticated \textit{ansatz} components for capturing electron correlation.
A key benefit to VMC is the ability to include intricate explicit descriptions of electron correlation through Jastrow factors, which add functional dependence on electron-nulcear, electron-electron, and higher order distances.
Benchmarking has shown simple wavefunctions comprised of a single Slater determinant and Jastrow factor can attain 3 kcal/mol accuracy\cite{petruzielo2012approaching_QMCChemicalAccuracy_Also_CommentOnCIParametersInPresenceOfJastrow_AFewIncrease}, while more sophisticated Jastrow functions can even capture strong correlation typically unattainable with a single Slater determinant\cite{van2017suppressing_Beatrice_NumberCounting_StrongCorrelation}.
In the exact same role that explicit correlation plays in F12/R12 theories\cite{kong2012explicitly_F12R12_Review,klopper2006r12_R12_Review_TenNo}, a well designed Jastrow factor can capture electron correlation beyond the basis set limit, mitigating the need to converge a CI expansion with respect to a long tail of small CI parameters and basis sets.

However, for successful QMC optimization within chemical accuracy, or for excited states or strongly correlated systems, a multi-Slater determinant expansion can be preferable.
Early works focused on CI expansions from complete active space (CAS) calculations. \cite{schautz2004optimized,petruzielo2012approaching_QMCChemicalAccuracy_Also_CommentOnCIParametersInPresenceOfJastrow_AFewIncrease,toulouse2007optimization_LM_C2,umrigar2007alleviation_LM,toulouse2008full,morales2012multideterminant}
However, these expansions are often truncated far from the inclusion of all determinants, typically including only a few hundred determinants or CSFs.
Two recent advancements have accelerated the inclusion of extended lists of optimizable CI paramters.
First, improvement of the sampling algorithm of Slater determinants via Sherman-Morrison updating\cite{nukala2009fast_FirstSM_Paper_I_Think,clark2011computing_SMMath_For_MultiSlater_Updates,scemama2016quantum_SMMath_for_CIPSI_DMC} as well as carrying derivatives through these updates\cite{filippi2016simple_MultiSlater_Derivative_Machinery,assaraf2017optimizing_Filippi_SMDerivativeMachinery} has allowed for significant gains in QMC optimization efficiency.
Second, a revitalization of sCI methods.
First pioneered by CIPSI\cite{huron1973iterative_OriginalCIPSI}, but with many groups recently developing and optimizing their own methods\cite{sharma2017semistochastic_DICE_1,chilkuri2021comparison_NEESE_Sci,tubman2020modern_ASCI_SCI,garniron2018selected_LOOS_SCI}, selected CI can approach full CI accuracy with a fraction of the combinatorally large number of determinants.
A common procedure has become utilizing sCI to generate lists of determinants which are then augmented with Jastrow factors to create the QMC trial wavefunction\cite{dash2018perturbatively_CIPSIVMC_GeometryOptimization,giner2013using_CIPSIForDMCFixedNodes_NoVMC,cuzzocrea2022reference_CNDyes,dash2021tailoring,dash2019excited_FilippiPT2SCI_ExcitationEnergies}.
Utilizing sCI as the foundation for multi-Slater wavefunctions maintains several advantages over CAS style wavefunctions, including lessening issues with choice of active orbitals, the ability to gradually increase the number of determinants by simply selecting more rather than having to expand and active space, and containing out of active space and out of valence excitations CAS would miss entirely.
It is worth mentioning sCI has proven useful not just in VMC, but also in DMC\cite{giner2013using_CIPSIForDMCFixedNodes_NoVMC,giner2015fixed_F2CIPSI_DMC,scemama2016quantum_SMMath_for_CIPSI_DMC,caffarel2016using_CIPSI_Nodes_For_DMC} and AFQMC\cite{mahajan2022selected_SCI_For_AFQMC}.

While sCI is an excellent source of Slater determinant expansions for VMC, the exact details of how to construct the trial wavefunction remain unclear.
First, while an sCI wavefunction is formally a truncation of the FCI wavefunction, even modest sized sCI expansions ($\approx 10^6-10^7$ determinants) are too large for any current VMC code to successfully optimize.
Therefore, sCI expansions must be further truncated before being passed into VMC.
Various schemes have been developed for how to appropriately size an sCI expansion for VMC optimization, based on matching quantities such as the sCI PT2 correction\cite{dash2019excited_FilippiPT2SCI_ExcitationEnergies} or the VMC variance\cite{robinson2017excitation_PJVarianceMatching,garner2020core}.

Both of theses schemes make decisions about which determinants to include
before accounting for the effects of the Jastrow factor.
This matters, because there is likely to be redundancy between Jastrow parameters and CI parameters, especially in the recovery of weak correlation.
One may expect that determinants included by sCI for the purpose of capturing weak correlation may, when re-optimized in the presence of a Jastrow factor, see their coefficients decrease as the correlation they are after has already been described at least in part by the Jastrow.
However, as has been noted elsewhere\cite{petruzielo2012approaching_QMCChemicalAccuracy_Also_CommentOnCIParametersInPresenceOfJastrow_AFewIncrease}, this is not always the case, and in some systems certain determinants see significant rises in their coefficients during VMC optimization.
Either way, it would not be surprising if the list of the ``most important''
determinants differs between VMC and sCI, or, more generally, differs with
and without the use of explicit correlation methods.
At present, however, it is not clear how best to determine the optimal
determinant inclusion order for multi-Slater Jastrow wavefunctions.

Let us now turn our attention to the challenges of wavefunction optimization
within VMC. 
The linear method (LM)
\cite{sorella2007weak_OriginalSR,nightingale2001optimization_OriginalLM}
is one of the most effective VMC optimizers, 
and since its introduction has continued to be improved,
\cite{neuscamman2012optimizing_Eric_OptimizingLargeParameterSets,sabzevari2020accelerated_SandeepLMAcceleration,zhao2017blocked_ChrisBLM,otis2019complementary_Leons_Optimizer}
both in terms of its efficacy at getting close to the true energy minimum
and in its computational efficiency.
Although the challenges of working with large parameter sets have been partially
mitigated by these developments, optimizing wavefunctions with large numbers
of variational parameters remains a challenge.
Two key issues underpin this challenge: (1) the memory cost associated with constructing matrices of size $N_p^2$, where $N_p$ is the number of variational parameters, and (2) the effects of statistical uncertainty
on gradient estimates and, in particular, on the nonlinear diagonalization of the LM eigenproblem when large numbers of variational
paramters are present. \cite{otis2022optimization_Leon_Stability_Paper}
The memory constraint can be lowered by avoiding
the explicit storage of the full Hamiltonian and overlap matrices.
\cite{neuscamman2012optimizing_Eric_OptimizingLargeParameterSets,zhao2017blocked_ChrisBLM,sabzevari2020accelerated_SandeepLMAcceleration}
Alternatively, there has been much recent interest in avoiding these matrices entirely via accelerated gradient descent methods,
\cite{schwarz2017projector_Descent_1,sabzevari2018improved_Sharma_Descent_2,mahajan2019symmetry_Sharma_Descent_3,luo2019backflow_Descent_4}
whose memory footprint grows only as $N_p$.
The other key issue --- the stochasticity of the detrivatives and matrix elements --- can be mitigated through modified estimators
\cite{umrigar2005energy_Modified_Estimator_1,toulouse2007optimization_LM_C2,pathak2020light_ImprovedGradientEstimators}
and step control measures
\cite{toulouse2008full,otis2022optimization_Leon_Stability_Paper}
akin to more general trust radius methods.
Despite these advances, it remains true that optimizing wavefunctions
with over a thousand parameters creates challenges in VMC, especially
if one desires to converge an optimization precisely with minimal
sampling effort.
As we will explore below, even highly effective optimizers can
benefit from excluding wave function parameters
that, at the chosen level of statistical sampling, turn out to
have statistically unresolvable effects.

With these optimization issues in mind, we return to the question of how to create a determinant expansion for a VMC trial wavefunction.
Because VMC is a variational method, optimization concerns aside, additional determinants and therefore variational flexibility should only help--and never hurt--the variational energy estimate.
However, with optimization difficulty increasing with the number of variational parameters, both in terms of memory economy and statistical precision, it may not be optimal, in a statistical sense, to expand the CI expansion too far.
In theory, the inclusion of a Jastrow factor should allow for shorter determinant lists--however there is no current \textit{a priori} way to know the optimal length of this list nor the determinants it should contain.
Here, we find that adding additional determinants not only slows optimization, but that it can adversely affect the variational energy achieved at a given sampling effort.
We present evidence suggesting that one of the causes of this
difficulty is simply that many parameters' potential benefits are
not statistically resolvable, such that the inclusion of these
parameters serves only to increase optimization noise.
Finding the ideal set of determinants at a given sample size therefore requires balancing the variational benefits of including more determinants with the statistical optimization challenge that doing so creates.
As the optimal balance is almost certainly dependent on sample size and the inherent statistical difficulty of the system being modeled, adaptive approaches are highly desirable.
We compare a few simple adaptive approaches to the naive inclusion of a large determinant list and find that adaptive strategies show promise in delivering better results with fewer determinants at reduced sampling efforts.

The remainder of the paper is organized as follows.
In Section \ref{sec:theory}, we provide an overview of VMC, trial wavefunction preparation, and the optimization methods we utilize.
In Section \ref{sec:methods} we introduce our test system and outline the computational procedure.
We provide evidence for our claims that optimization quality is dependent upon the optimizer's ability to statistically resolve parameters in \ref{sec:Sampling} and \ref{sec:ResolvingCI}.
Finally, in sections \ref{sec:MedianSCI}, \ref{sec:ParamSCI}, and  \ref{sec:TrueSCI}, we provide details of sCI inspired methods for properly choosing a CI expansions that only include statistically resolvable parameters.

\section{Theory}
\label{sec:theory}
\subsection{VMC \& Trial Wavefunction}
\label{sec:VMC}
In VMC, the central quantity of interest is the variational energy of a trial wavefunction
\begin{equation}
    E_V=\frac{\bra{\Psi_T}\hat{H}\ket{\Psi_T}}{\braket{\Psi_T}{\Psi_T}} \geq E_0
\end{equation}
which can be estimated via real space sampling 
\begin{equation}
E_V\approx\frac{1}{N}\sum^N_iE_L(\mathbf{R}_i)
\end{equation}
where $E_L$ is the local energy ($E_L=\frac{\hat{H}\Psi_T}{\Psi_T}$) and $\mathbf{R_i}$ are real space electron coordinates sampled via Markov chain Monte Carlo, typically sampled via Metropolis sampling the $|\Psi_T|^2$ distribution\cite{metropolis1953equation}.
Being a variational theory, the optimal variational energy estimated in a VMC calculation is strictly greater than or equal to (within statistics) the true ground state energy, $E_0$.
As a consequence, increasing the \textit{ansatz} complexity, and therefore variational flexibility, should strictly lower the variational energy.

For molecular calculations, a common form of trial wavefunction, \iket{\Psi_T}, is a multi-Slater Jastrow with functional form
\begin{equation}
    \ket{\Psi_T}=e^J \sum_Ic_I\ket{I}
\end{equation}
where $e^J$ is the Jastrow factor and \iket{I} is an anti-symmetric Slater determinant with associated coefficient $c_I$.
Jastrow factors are totally symmetric with respect to electron exchange, and commonly include one and two body components, which explicitly depend on electron-nucleus and electron-electron distances, respectively.
The multi-Slater component ensures Fermionic antisymmetry with respect to particle exchange.
The individual determinants are composed of electrons in single particle orbitals which can be obtained from standard quantum chemical methods.

The simplest version of the Fermionic component of the wavefunction is a single determinant, typically the Aufbau Hatree-Fock determinant.
However, VMC benefits from the same machinery that composes traditional CI theories--particularly for capturing strong electron correlation.
For example, VMC utilizing only the Hartree-Fock determinant retains some incorrect ionic dissociation character, while stating from a CASSCF wavefunction give the qualititatively correct neutral dissociation\cite{toulouse2008full}.
For the reasons we discussed in the introduction, an excellent choice for constructing a CI expansion is sCI. Indeed, beyond VMC, this approach
to trial functions has also seen significant use in QMC more broadly.
\cite{giner2013using_CIPSIForDMCFixedNodes_NoVMC,giner2015fixed_F2CIPSI_DMC,scemama2016quantum_SMMath_for_CIPSI_DMC,caffarel2016using_CIPSI_Nodes_For_DMC,mahajan2022selected_SCI_For_AFQMC}

Briefly, sCI aims to approximate the FCI wavefunction by constructing a small wavefunction which has "selected" only the most important determinants.
Each sCI iteration consists of two main stages.
First, the Hamiltonian is diagonalized in the current variational subspace, \iket{\Psi^{(i)}}.
Second, determinants not currently in the wavefunction are "selected" for inclusion in the next variational subspace \iket{\Psi^{(i+1)}}.
The exact details of how the selection scheme is done depends on the sCI implementation, however most utilize perturbative arguments to ranks a determinant's importance based on it's Hamiltonian matrix element with determinants currently in the wavefunction, that is
\begin{equation}
    \texttt{Imporance of }\ket{I}\propto \left|\bra{I}\hat{H}\ket{J}\right|
    \left\{ \begin{array}{r} 
        \ket{I}\notin\ket{\Psi^{(i)}} \\
        \ket{J}\in\ket{\Psi^{(i)}} \\
        \end{array} \right.
\end{equation}
This procedure can be repeated until sufficient convergence is reached.
Practically, however, many implementations rely upon a PT2 correction after growing a sufficiently large variational space.
In this work, we use a CASCI wavefunction as our reference, which can be viewed as a completely converged variational (ie, no PT2) sCI calcualtion in an active space.
While wildly successful in recently years, sCI retains the same formal combinatorial scaling as FCI.
Further, capturing the sharp features of electron correlation require converging long, slowly decaying tails sCI wavefunctions in large basis sets.

Preparing a sCI wavefunction for VMC optimization involves first augmenting the wavefunction with appropriate Jastrow factors.
A decision must then be made to length of CI expansion to include.
The guiding principles for this decision are based upon balancing retaining the sCI accuracy while minimizing the number of parameters needing optimized.
Different schemes based on either sequentially expanding an sCI calculation or truncating a much larger calculation exist\cite{dash2021tailoring,dash2019excited_FilippiPT2SCI_ExcitationEnergies,robinson2017excitation_PJVarianceMatching,garner2020core} but we will look in Section \ref{sec:EasingOpt} to make better decisions informed by the statistical resolution of CI parameters.

\subsection{VMC Optimizers}
\label{sec:OptimizationTheory}
The goal of a VMC optimization is to find the ideal parameter set which minimizes a target quantity, often the variational energy or its variance.
For a multi-Slater Jastrow wavefunction, a typical set of optimizable parameters is composed of the CI expansion coefficients, orbital optimization parameters, and Jastrow parameters.
Ideally, these parameter sets capture strong electron correlation, oribital relaxation, and weak correlation, respectively.
Realistically, however, it is not possible to decoupled the effects of each of these components.
For example, single excitations in a CI expansion can capture some, but not all orbital relaxation, while double excitations can provide some details of weak correlation.
The goal of a VMC optimizer is to efficiently balance these competing effects to reach the optimal parameter set when given statistically noisy estimates of the target function and required gradients. 

As discussed above, the LM has been highly successful in VMC optimization.
This method Taylor expands the wavefunction to first order in its variational parameters.
\begin{equation}
    \ket{\Psi}=\sum_{j=0} c_j \ket{\Psi^j}
\end{equation}
Here $\ket{\Psi^0}\equiv\ket{\Psi_T}$, is the wavefunction itself,
while \iket{\Psi^j} for $j>0$ is the derivative of the wavefunction
with respect to variational parameter $j$.
With this expansion, the optimal parameter update direction can be obtained via solving the generalized eigenvalue equation
\begin{equation}
    \mathbf{H}\mathbf{c} = E \mathbf{S}\mathbf{c}
\end{equation}
The elements of $\mathbf{H}$ and $\mathbf{S}$ are given by
\begin{equation}
    H_{ij} = \dfrac{\bra{\Psi^i}\hat{H}{\ket{\Psi^j}}}{\braket{\Psi}{\Psi}}
\end{equation}
\begin{equation}
        S_{ij} = \dfrac{\braket{\Psi^i}{\Psi^j}}{\braket{\Psi}{\Psi}}
\end{equation}
In a parallel with sCI, if $i,k$ are both CI parameters, we interpret $H_{ik}$ as the matrix element between determinants \iket{I} and \iket{K}, that is
\begin{equation}
    H_{ik}=\frac{\bra{I}e^J\hat{H}e^J\ket{K}}{\braket{\Psi}{\Psi}}
\end{equation}
although in this case the matrix element involves explicit correlation via the Jastrow factor, $e^J$.
Further, the first row (or column) of $\mathbf{H}$ can be interpreted as the matrix element between any given determinant and the total current wavefunction,
\begin{equation}
    H_{i0}=\frac{\bra{I}e^J\hat{H}\ket{\Psi_0}}{\braket{\Psi}{\Psi}} = \frac{\bra{I}e^J\hat{H}e^J\sum_kc_k\ket{K}}{\braket{\Psi}{\Psi}}
\end{equation}

The linear method is a powerful tool, but is subject to two main drawbacks.
First, the required Hamiltonian and overlap matrices are subject to the stochastic noise inherent to the sampling of it's constituent components.
To attempt to reduce statistical noise, multiple versions of modified estimators have been proposed and successfully utilized\cite{umrigar2005energy_Modified_Estimator_1,toulouse2007optimization_LM_C2}.
A recent approach by Ammar, Giner and Schemama\cite{ammar2022optimization_Hybrid_CI_with_Transcorrelation} removes optimization of CI parameters from VMC altogether, instead diagonalizing a more traditional CI problem with additional transcorrelated matrix elements stochastically sampled.
A significant effort in this work focused on creating appropriate modified estiamtors to reduce the statistical noise in the sampled quantities, underscoring the difficulty stochatic noise present any optimization scheme.

The second major drawback of the linear method is the memory cost associated with constructing and storing matrices of size $N_p^2$, where $N_p$ is the number of optimizable parameters.
Ways to avoid the explicit storage of full sized matrices\cite{neuscamman2012optimizing_Eric_OptimizingLargeParameterSets} have been introduced to help circumvent the memory bottleneck.
In particular, consider the blocked linear method\cite{zhao2017blocked_ChrisBLM}, which subdivides the entire linear method matrix into smaller sub blocks of parameter sets, which are each individually diagonalized.
Each block determines an appropriate update direction, and the blocks are recombined taking into account inter-block coupling to produce the overall parameter update.
Dividing the linear method matrix in this way reduces the memory footprint from $N_p^2$ to $N_p^2/N_b$, where $N_b$ is the number of blocks, all at very little penalty to the overall optimization progress\cite{otis2019complementary_Leons_Optimizer}.

An alternate approach to address the memory bottleneck of the traditional linear method is by pursuing gradient descent approaches\cite{schwarz2017projector_Descent_1,sabzevari2018improved_Sharma_Descent_2,mahajan2019symmetry_Sharma_Descent_3,luo2019backflow_Descent_4}.
In descent methods, no explicit matrices are constructed, reducing the memory cost to just that of storing the gradient information on each individual parameter value, or $N_p$.
In the simplest form, one can take a steepest descent approach by updating each individual parameter by taking a step in the direction of and proportional to it's current gradient values.
While functionally steepest descent can converge to a minimum, accelerated descent methods, which take into account the optimization history in some form, can converge much more quickly in practice.
\cite{schwarz2017projector_Descent_1,sabzevari2018improved_Sharma_Descent_2,mahajan2019symmetry_Sharma_Descent_3,luo2019backflow_Descent_4,kingma2014adam_ADAM_Optimizer}

Here, we will use a hybrid approach to optimization, which utilizes primarily relatively low cost iterations of descent along with periodic block linear method updates.
The details of this optimizer are described elsewhere.
\cite{otis2019complementary_Leons_Optimizer}
Briefly, the main optimization procedure alternates sections of accelerated descent with a few iterations of block linear method optimization.
This main optimization is expected to draw close to the optimal parameter set, from which we increase our sampling effort and perform a long section of accelerated descent finalization.
These iterations can assist in more tightly converging to the true minimum as well as serve as a convenient way to collect optimization statistics.

\section{Methods}
\label{sec:methods}
For all calculations, we have used a water molecule with one of the OH bonds stretched to about $\approx1.4$ times it's equilibrium distance.
The exact geometry is given in Section \ref{sec:WaterGeom}, Table \ref{tab:water_geom}.
Hartree Fock orbitals are used the ccECP-cc-pVDZ pseudopotential basis\cite{bennett2017new_pseudopotential_basis}.
An (8e,8o) CASCI is performed to generate a multi-determinant expansion utilizing DICE\cite{sharma2017semistochastic_DICE_1,holmes2016heat_DICE_2} (with $\epsilon_1$=0) in conjuction with PySCF\cite{sun2020recent_PySCF,smith2017cheap_DICE_PySCF}.
Symmetry is used for generating the HF orbitals, but turned off for the CASCI calculation.
We remove symmetry disallowed determinants from the CASCI expansion by removing any determinants with coefficients with magnitude $<10^{-8}$, resulting in a CASCI space containing 2448 determinants.
The multi-Slater wavefunction is augmented with one and two body Jastrow factors.
Each atom has its own unique one body Jastrow.
All Jastrows are composed of cardinal-cubic B-Splines\cite{kim2018qmcpack_QMCPACK} with 10 spline points evenly distributed to a cutoff radius of 10 bohr.

All VMC optimizations are energy minimizations with samples drawn from $|\Psi|^2$. 
Unless noted otherwise, all optimizations consist of 16 macro iterations of alternating periods of accelerated gradient descent and block linear method followed by 1000 iterations of accelerated descent finalization.
The relevant amounts of sampling effort for this optimization procedure, which we will call our standard optimizer, are listed in Table \ref{tab:optimizer_sampling_effort}.
All CI and Jastrow parameters are optimized simultaneously in all optimizations.
The orbitals are are held fixed and not optimized, both for simplicity in this preliminary study of the optimization benefits of
within-VMC determinant selection, and because
the availability of state-specific orbitals from CASSCF often
makes further orbital optimization within VMC unnecessary.
\cite{otis2021combining_Leon_SSCAS_QMC}
Energies are reported as the average and standard error over the final 500 iterations of accelerated descent finalization\cite{otis2019complementary_Leons_Optimizer}.
Parameter values reported are the parameters value averaged over the same 500 iterations.
All VMC calculations are performed in a development version of QMCPACK\cite{kim2018qmcpack_QMCPACK}.

\begin{table*}[hptb]
    \centering
    \begin{tabular}{|l|c|r|r|r|}
        \hline
        Optimizer & Macro iterations & Optimization Method  & Micro iterations &
        \hspace{6mm} Samples \hspace{6mm}  \\
        \hline
        \hline
        \multirow{2}{*}{Hybrid Method} & \multirow{2}{*}{16} & Accelerated Descent & 100 & 30,000$~$/$~$micro \\
        &&Block Linear Method & 3 & 500,000$~$/$~$micro \\
        \hline
        Descent Finalization & 1 & Accelerated Descent & 1000 & 100,000$~$/$~$micro \\
        \hline
        \multicolumn{5}{|r|}{Total Sampling Effort: \hspace{2.4mm}
        172,000,000 \hspace{1mm}} \\
        \hline
    \end{tabular}
    \caption{Details of the standard VMC optimization used in this work.}
    \label{tab:optimizer_sampling_effort}
\end{table*}

\section{Results \& Discussion}

\subsection{Effect of Sampling on VMC Optimization}
\label{sec:Sampling}

    \begin{figure*}
        \centering
        \includegraphics[width=\textwidth]{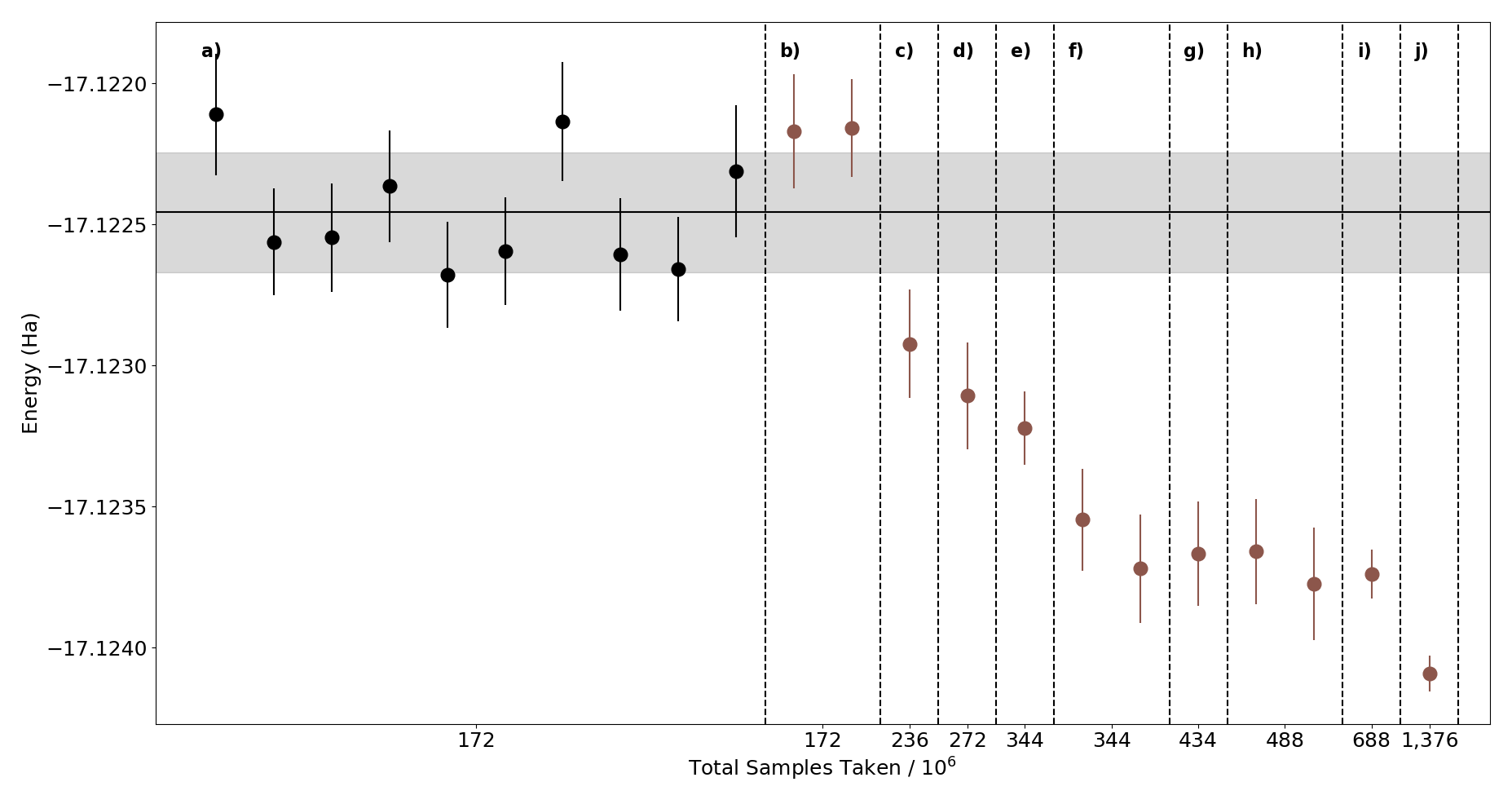}
        \caption{VMC energy for stretched water.  
        All points are the same Jastrow-CASCI wavefunction.  
        Different sections divided by dashed lines are different optimization procedures, ordered left to right in by increasing total number of samples taken.
        Section \textbf{a)} is 10 independent optimizations using our default optimizer, the sampling effort of which can be found in Table \ref{tab:optimizer_sampling_effort}.
        The solid black horizontal line is the average of these 10 optimizations.
        Sections \textbf{e), i),} and \textbf{j)} utilize the same number of macro iterations as \textbf{a)}, but with 2x, 4x, and 8x the number of samples per iteration, respectively.
        The details of the remaining optimizations can be found in Table \ref{tab:all_optimizers_fig_1}.}
        \label{fig:10TrialData}
    \end{figure*}

We begin with optimizing the multi-Slater Jastrow wavefunction including the full CASCI list of determinants, which contains 2448 determinants (2447 optimizable CI parameters).
The optimization was run 10 times with different random seeds, each time with the CI parameters initialized to their CASCI values.
The energy of each of these optimizations, as well as their collective average and standard deviation, are shown in black in section \textbf{a)} of Figure \ref{fig:10TrialData}.

The brown points in sections \textbf{b)-j)} of Figure \ref{fig:10TrialData} all contain the same complete Jastrow-CASCI wavefunction, but optimized with different details of optimization.
Each point is an individual optimization starting from CASCI coefficients with a different random seed, with the total number of samples over the course of optimization noted on the horizontal axis.
Sections \textbf{e), i),} and \textbf{j)} all were optimized with the standard optimizer scheme (as presented in Table \ref{tab:optimizer_sampling_effort}), but with 2x, 4x, and 8x samples at each individual iteration, respectively.
The remaining sections all shift the distribution of samples between optimization steps, for example panel \textbf{b)} runs half the total number of optimization iterations (8 hybrid method, 500 descent finalization), but with 2x samples at each individual iteration.
The details of the how samples are distributed over iterations for the remaining sections are presented in Section \ref{sec:Fig1OptimizationDetailsTable}, Table \ref{tab:all_optimizers_fig_1}.

Reading these sections from left to right, each optimizer contains a monotonically increasing amount of total samples.
The resulting VMC energies show a monotonic decrease with increasing sampling effort.
Further, the optimized VMC energy seems to only depend on total number of samples and not how these samples are distributed during the different stages of optimization.
Using section \textbf{b)} as an example, the same total number of samples distributed among half as many optimization iterations produces statistically indistinguishable energies from the average from the points in section \textbf{a)}.
At the two furthest right points, which again were optimized with the standard optimization scheme with 4x and 8x as many samples, respectively, the energy continues to decrease and does not
yet appear to have converged with respect to increasing sampling effort.

The consistency among the first ten optimizations implies that
the optimizations find the same minima.
We also see that, when using the standard optimizer and repeating the optimization ten times over, the optimization uncertainty is about on par with the energy uncertainty of each individual optimization's final wavefunction.
The steady decrease of the energy with respect to sample size strongly suggests that more samples are simply tightening our statistical certainty around the same minima, not finding a new minima.
In our most aggressive optimization, which simply took 8x as many samples during each optimization step as the reference ten optimizations, the VMC energy lowers from -17.1225(2) to -17.12409(6).
It is worth noting that this energy lowering is about 1.5 $\mathrm{mE_h}$ and is larger the oft-cited "chemical accuracy" of 1 kcal/mol.
What components of the wavefunction and its parameterization are driving the uncertainty that appears to limit the optimization at smaller sample sizes, and what precisely is it about increased sampling that overcomes this difficulty?

\subsection{Statistically Resolving CI Parameters}
\label{sec:ResolvingCI}

\begin{figure*}
    \centering
    \includegraphics[width=\textwidth]{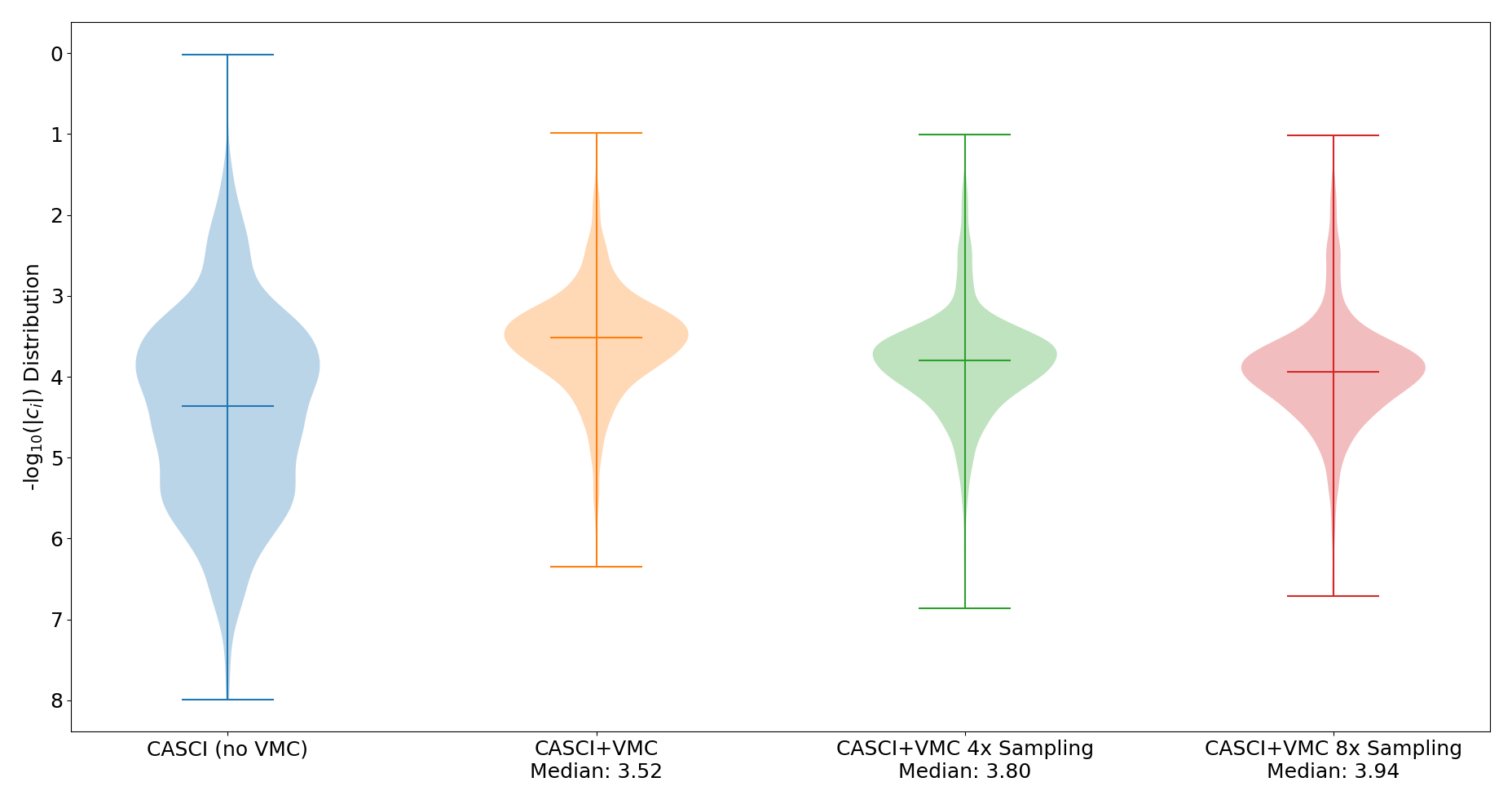}
    \caption{Violin plot distributions showing the logarithmic spread of CI coefficients.  
    The CASCI wavefunction without VMC optimization is shown at left.  
    The three distributions on the right are VMC-optimized wavefunctions containing all the
    CASCI determinants, with the optimizations performed at three different levels of sampling effort.}
    \label{fig:ViolinData}
\end{figure*}

To investigate the source of the sample-size-dependent optimization quality, we'll look at the distributions of CI parameters, as plotted in Figure \ref{fig:ViolinData}.
The violin plots are of the $-\log_{10}$ of the CI parameters.
The blue, left-most distribution shows the CI parameters at the CASCI level.
As could be expected, there are very few parameters with magnitude $>10^{-1}$, namely two determinants which are the Aufbau and strongly correlating $(\sigma)^2\rightarrow(\sigma^*)^2$ determinant breaking the stretched OH bond.
The remaining determinants are distributed non-uniformly across about 5 orders of magnitude.

The next distribution in Figure \ref{fig:ViolinData} shows the CI parameter distribution
after VMC optimization with our standard scheme (Table \ref{tab:optimizer_sampling_effort}).
It is worth noting how different this distribution appears compared to the CASCI distribution.
While the CASCI distribution is non-uniformely distributed across several orders of magnitude, once optimized via VMC, the distribution becomes much more sharply peaked in what appears to be a single distribution.
The median of all parameters is used as a rough measure for the center of this distribution, located at 3.52 (or raw CI value of $\approx 3 \times 10^{-4}$).
Of course, a significant change in the CI variables is expected when reoptimizing them
in the presence of the Jastrow.
However, it is not at all clear that the optimal with-Jastrow distribution of CI parameters should look like what we see in Figure \ref{fig:ViolinData}, with the bulk of the
distribution peaked at a higher value than in raw CASCI and with no very small
coefficients.
Indeed, we believe that the distribution we actually observe after VMC optimization is
shaped at least as much by the optimization's limited statistical resolution as by
the physics of the electronic structure.
Looking again at the median of the distribution, we suspect that its position is
a rough guide to our optimization's ability to statistically resolve CI parameters.
We are doubtful that CI parameters with true optimal values near or below this median
can be distinguished from statistical noise at the sample size employed.

To check if this distribution truly is just the distribution of statistical noise on parameter values, we plotted the same distributions for the same optimizer but with 4x and 8x times as many samples per iterations (sections \textbf{i)} and  \textbf{j)} from Figure \ref{fig:10TrialData}).
These distributions appear to be shifted toward smaller overall CI values (larger $-\log_{10}$ value), which is consistent with the understanding that the statistical noise decreases with increasing sampling.
Further, with a 4 fold increase in sampling, our statistical resolution should increase by a factor of 2.
If this distribution was purely stochastic noise controlled by normal statistics, the median should increase by a shift of $\log_{10}2\approx0.30$, and we see an increase from 3.52 to 3.80, or 0.28.
A similar analysis for 8x sampling effort would increase the mean by $\log_{10}\sqrt{8}\approx0.45$, and we see an increase from 3.52 to 3.94, or 0.42.
While this analysis is not a conclusive proof that the majority of the CI distribution is purely statistical noise, the remarkable trend along with increased sampling certainly supports this perspective. 

Returning to the median of the CI parameters optimized with our standard sampling effort, 3.52, it should be noted that a vast majority of CI parameters from the CASCI distribution are smaller in magnitude than this median (the median of the CASCI distribution is 4.36!).
While again it is not surprising that the CI parameters are significantly different when optimized in the presence of a Jastrow, having the average CI parameter raise it's value by an order of magnitude is not expected.
If anything, one would expect weak correlation recovery by the Jastrow factor to decrease the importance of minor determinants. 
Critically, we believe that it is this inability to statistically resolve a significant portion of CI expansion that limits our VMC optimizations with smaller sample sizes.
These unresolved parameters may serve only to add noise to the wavefunction and the optimization, both of which are expected to raise the energy.
This conclusion is supported by a decrease in variational energy as sampling is increased--higher sampling helps statistically resolve more CI parameters, helping to mitigate the harm done by these noisy parameters.
What if, instead, we selectively removed the determinants that are lost in this noise?

\subsection{Easing Optimization}
\label{sec:EasingOpt}
One approach to mitigating the optimization difficulty associated with
statistically unresolvable CI parameters is to simply converge calculations
with respect to increased sample sizes.
Increased sampling can quickly become impractical, however, and pushing
past 1.3 billion total samples in the optimization for a single water molecule
does not bode well for the wider world of chemistry.
We therefore look to develop strategies for the identification and
removal of unhelpful parameters.

First, we will study two methods that start from a large set of parameters
and, after some initial optimization, decide which are worth keeping
and which are better to set aside.
Then, we will study a method that starts small with just the Jastrow and
the Aufbau determinant and then pulls in additional determinants based on
statistical estimates of their matrix elements with the existing wavefunction.
While this third approach is, at least in its current form, too crude to be
fairly described as a sCI in the tradition sense, it does share the same
basic outline, and our data suggests that in future it may be beneficial
from the standpoint of optimization efficiency to more fully import
the machinery of sCI into VMC.

\subsubsection{Median Removal}
\label{sec:MedianSCI}

\begin{figure*}
    \centering
    \begin{subfigure}{\textwidth}
        \includegraphics[width=\textwidth]{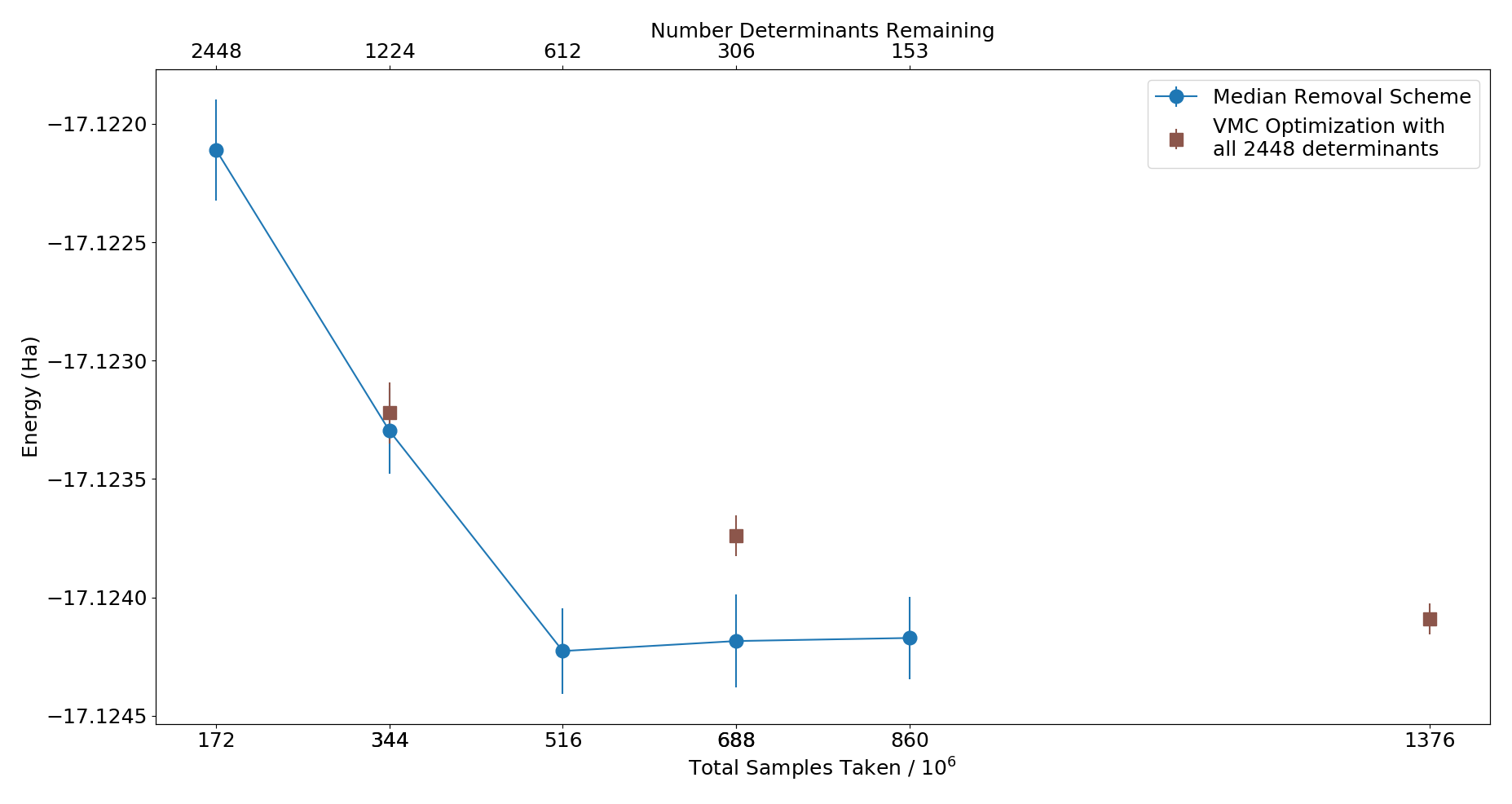}
        \caption{VMC energy for median removal scheme (see text).
        The connected blue circles show the optimization progress.
        Each blue point contains half as many determinants as the point previous to it as indicated on the upper horizontal axis, and is fully reoptimized with our standard optimizer.
        For comparison, the brown squares are the full Jastrow-CASCI wavefunction optimized in our standard optimizer with 2x, 4x, and 8x sampling as read left to right (these are identical to the points in Sections \textbf{e), i),} and \textbf{j)} of Figure \ref{fig:10TrialData}).}
        \label{fig:Median_Removal_E}
    \end{subfigure}
    \hfill
    \begin{subfigure}{\textwidth}
        \includegraphics[width=\textwidth]{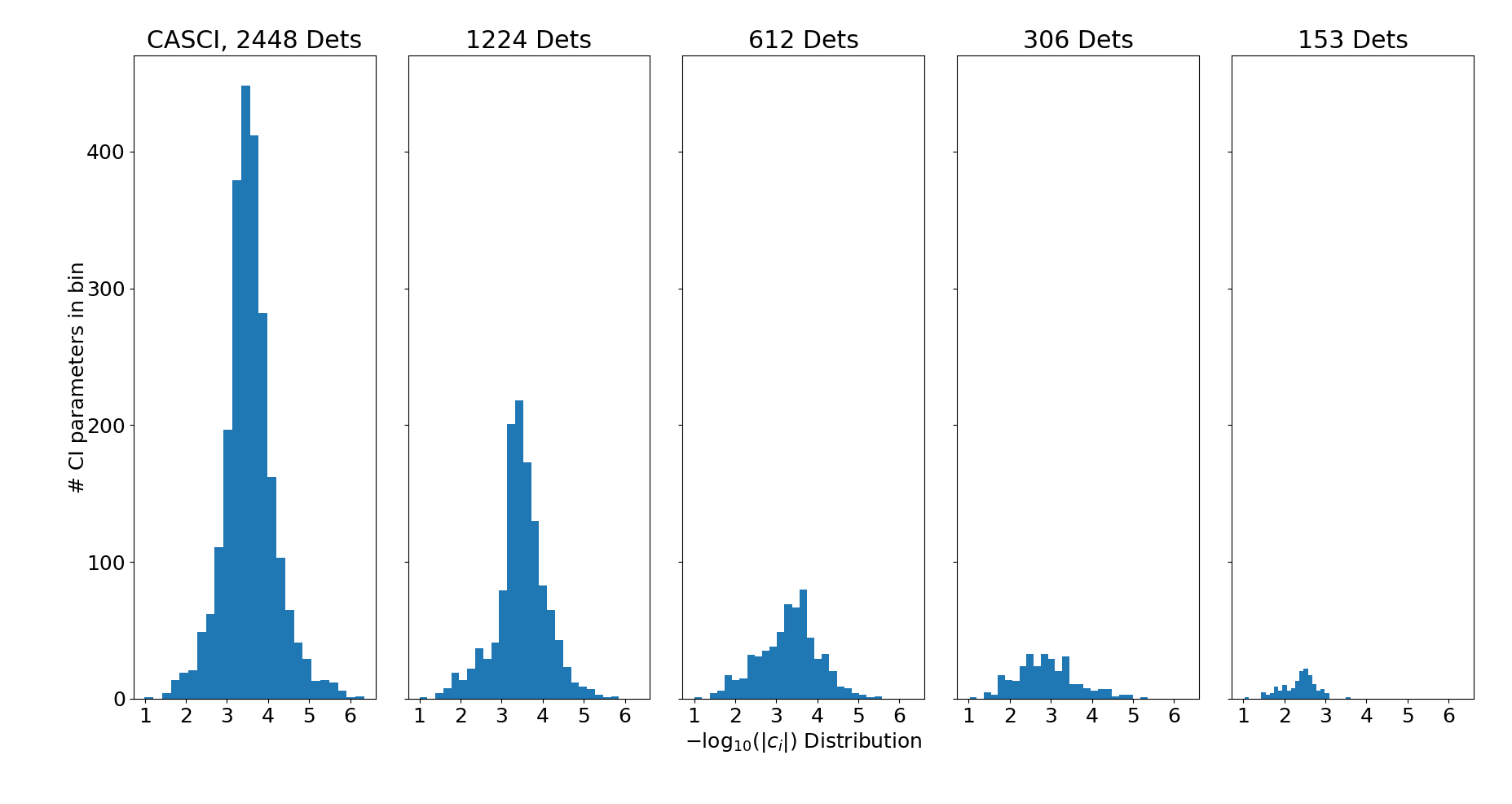}
        \caption{Histograms of orders of magnitude of CI parameters in median removal scheme wavefunctions.
        The histograms are for the blue circle wavefunctions with plotted in Figure \ref{fig:Median_Removal_E}.
        The leftmost histogram is identical to that of the orange CASCI wavefunction (with median 3.52) in Figure \ref{fig:ViolinData}.
        The histograms share common horizontal and vertical axes.}
        \label{fig:Median_Removal_Distributions}
    \end{subfigure}
    \caption{Median removal scheme}
    \label{fig:Median_Removal}
\end{figure*}

With the medians determined, a very simple approach is to remove all parameters with values suspected to be part of the statistical noise distribution.
To do this conservatively, after first optimizing the full Jastrow-CASCI wavefunction, we simply delete all CI parameters with values smaller than the median CI parameter value, then reoptimize through the standard optimizer.
It is important to say that while each wavefunction has gone through the standard optimization multiple times, because the memory of accelerated descent is not carried forward, each optimization is independent from the optimization that preceded it, and so likely less effective than one long optimization would be.
Starting from the wavefunction which produced the orange distribution (median 3.52) in Figure \ref{fig:ViolinData}, we ran and repeated this procedure 4 times, the results of which are shown in Figure \ref{fig:Median_Removal}.

The VMC energies of the wavefunctions optimized in this way are shown in Figure \ref{fig:Median_Removal_E}, along with the energies of the 2x, 4x, and 8x sampling energies from Figure \ref{fig:10TrialData} (sections\textbf{ e), i)} and \textbf{j)}, for reference).
The first time the wavefunction CI expansion is halved, the variational energy decreases by roughly a milihartree, from -17.1221(2) to -17.1233(2).
Upon the next halving, the energy again decreases, to -17.1242(2), which is in agreement with the most aggressive optimization achieved of the full CASCI wavefunction.  
The final two iterations don't statistically change the variational energy.

To further analyze these wavefunctions, we have plotted the $-\log_{10}$ CI coefficient distributions for each of the wavefunctions as histograms in Figure \ref{fig:Median_Removal_Distributions}.
We have kept the vertical axis the same for each distribution to emphasize the significant shrinking of the determinant list in each of the wavefunctions.
The first distribution is identical to the orange distrubtion (median 3.52) in Figure \ref{fig:ViolinData}, but now plotted as a raw histogram as opposed to a violin density plot.
As expressed in the discussion of Figure \ref{fig:ViolinData}, we believe the majority of the main peaked feature of this distribution (between about 3-4 $-\log_{10}$ in value) to be comprised of CI parameters who's true value is hidden by the optimization statistical noise.
After the first iteration of deleting half of and reoptimizing the CI parameters, the distribution remains mostly peaked around the same 3-4 $-\log_{10}$ value, but appears to be more skewed toward larger CI values as opposed to smaller.
Interestingly, by the second iteration of this scheme (612 Dets), when our variational energy has converged, there still seems to be significant number of determinants peak in the distribution in the same 3-4 regime.  
That peak is for the most part removed in the following iteration, and almost fully removed in the final iteration.
There are a number of possible interpretations for this.
First, when the parameter set is reduced, even at a fixed sample size, better steps are possible because the overall parameterization statistical noise has been reduced, as has been seen elsewhere\cite{otis2022optimization_Leon_Stability_Paper}.
With improved optimization steps, we may have greater statistical resolution on each CI parameter even at fixed sample sizes.
Second, the fact that the final three optimizations all produced statistically indistinguishable energies perhaps suggests that having some statistically unresolvable parameters (as would seem to be the case in the 612 Dets wavefunction), but not too many (as would be the case in the 1224 Det or 2448 Det case) can still lead to a statistically well resolved wavefunction.
Third, potentially even though we believe the majority of the peaked distribution in the CASCI optimization is composed of CI parameters we are not statistically resolving, possibly some determinants in this regime are in fact well statistically resolved, but have an overall small effect on the VMC energy.
It may be interesting to attempt to distinguish between these possibilities in future work.

\subsubsection{Parameter Statistical Removal}
\label{sec:ParamSCI}
\begin{figure*}
    \centering
    \includegraphics[width=\textwidth]{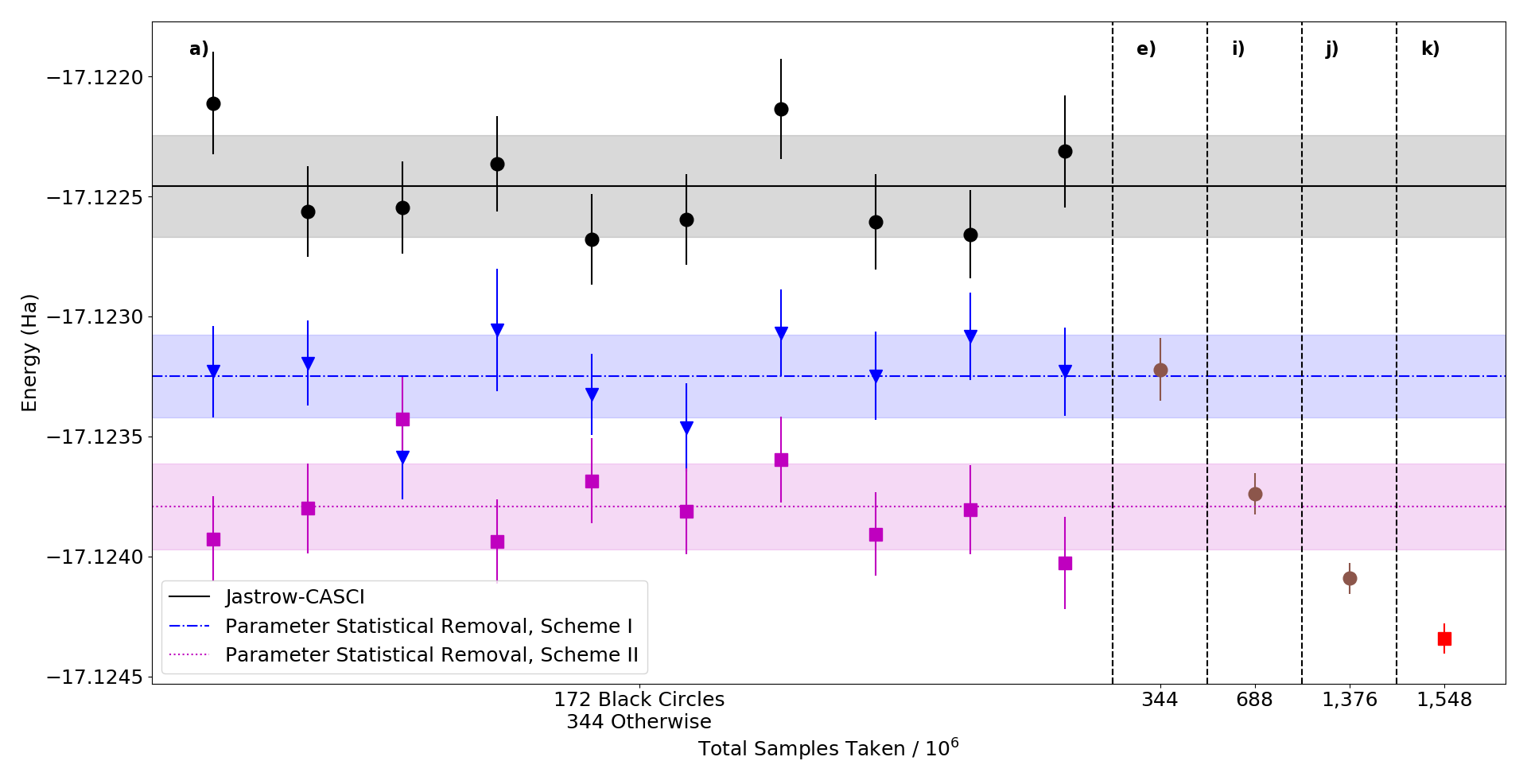}
    \caption{
    VMC Energies for the parameter statistical removal scheme (see text).
    The black circles in panel \textbf{a)} are identical to those in Figure \ref{fig:10TrialData}.
    Each blue triangle is the re-optimized (via the standard optimizer) VMC energy of a wavefunction with a CI expansion length selected via applying scheme I to the wavefunction which produced the black circle directly above.
    Similarly, the magenta squares are VMC energies but with CI expansion length selected via scheme II to the black circle above.
    For reference, panels \textbf{e), i)} and \textbf{j)} from Figure \ref{fig:10TrialData} are included, which contain the complete Jastrow-CASCI wavefunction optimized with the standard optimizer with 2x, 4x, and 8x as much sampling, respectively.
    Panel \textbf{k)} takes the furthest left magenta square and optimizes with the standard optimizer with 8x sampling.
    }
    \label{fig:RRScheme}
\end{figure*}

While the previous scheme removed parameters based on their standing in the an overall distribution of all current CI variables,  we now look for a way to flag and remove individual CI variables, independent of the overall distribution.
In this pursuit, we again key in on the idea that there is a fixed lower bound for what is a statistically resolvable CI parameter at a given sample size.
If the true value for a CI parameter is smaller than this lower bound, it can be expected that over the course of optimization these parameters update values will behave as statistical fluctuations around zero.

With this in mind, we applied the following protocol for removing individual CI parameters.
Utilizing the last 500 iterations of accelerated descent, we take the average and standard deviation for each individual CI parameter.
Should the standard deviation of an individual parameter's value be greater in magnitude than the parameter's average value itself, we deem that parameter to be statistically indistinguishable from 0, and delete it from our wavefunction.
We perform this check on all CI parameters and then repeat the standard optimization procedure.
We call this scheme Parameter Statistical Removal, Scheme I.
Looking to be extra aggressive in our parameter removal, we also define Scheme II, in which we add one additional condition--should any CI parameter change sign over the final 500 iterations of accelerated descent, we will consider that a sign the parameter is not statistically resolved and remove it as well.
This sign-change criterea is in addition to the criterea used in Scheme I.

The results of each of thsese schemes are shown in Figure \ref{fig:RRScheme}, with Scheme I in blue triangles and Scheme II in magenta squares.
To seed each of the removal schemes, we used one of the ten wavefunctions from section \textbf{a)} of Figure \ref{fig:10TrialData}, which are plotted here again vertically aligned with the wavefunctions they seeded.
The 2x, 4x, and 8x sampling Jastrow-CASCI optimization from Figure \ref{fig:10TrialData} (sections\textbf{ e), i)} and \textbf{j)}) are
again plotted for reference.

Both schemes produce lower energies than the
standard optimizer did when it was straightforwardly
applied to the full Jastrow-CASCI wavefunction.
Scheme I produces an energy comparable to when the
the standard approach is given 2x its normal sample size.
Again, because the accelerated descent memory is not carried forward, the reoptimized wavefunctions are not beneifiting from twice as much optimization as section \textbf{e)} would. 
Becuase these wavefunctions are each seeded by a different full Jastrow-CASCI optimization, they contain a range of total number of determinants, from 1285 to 1392, with an average of about 1361 determinants.
Comparing to the first median removal iteration, when the wavefunction has been cut to 1224 determinants, these optimizations produce statistically indistinguishable energies: -17.1232(2) for the average of the Scheme I trials
vs.\ -17.1233(2) for the 1224 determinant median removal wavefunction.
Both of these wavefunctions are likely still suffering from statistically unresolvable CI parameters, as evidenced by the even lower energies
achieved by later iterations of the median removal scheme.

Scheme II further improves the energy, decreasing to an average value of -17.1238(2), which is on par with the Jastrow-CASCI wavefunction optimized with 4x sampling effort (section \textbf{i)}), with a few optimizations overlapping with the 8x sampling (section \textbf{j)}) optimization within error bars.
The Scheme II wavefunctions contain between 587 and 715 determinants, averaging about 660 determinants.
Out of curiosity, we reoptimized one of these wavefunctions
(the one which produced the left-most magenta square)
with the standard optimizer with 8x sampling.
The resulting energy, -17.12434(6), is plotted in section \textbf{k)} in red
and is the lowest energy produced by any of the calculations in this study.
Importantly, this is lower in energy than the full Jastrow-CASCI wavefunction
optimized with a similar amount of sampling (section \textbf{j)}).
Thus, even at larger sample sizes, a smaller and more statistically
significant list of CI parameters leads to a better optimization
result compared to the full CASCI list.

At this point, it is worth considering computational efficiency
in greater detail.
Both median removal and parameter statistical removal rely upon
first running a VMC optimization on a large list of determinants,
and then removing parameters that appear to be impeding the optimization
and then reoptimizing. 
During Markov chain propagation, the value of each determinant
must be computed during each electron move.
Even with clever Shermann-Morrison approaches that mitigate this cost,
\cite{nukala2009fast_FirstSM_Paper_I_Think,clark2011computing_SMMath_For_MultiSlater_Updates}
evaluating a long list of determinants is more expensive than evaluating
a short list.
So, reducing the number of CI parameters reduces the cost of sampling and
thus optimization.
Performing two optimizations in this way --- first for the full Jastrow-CASCI 
wavefunction and then for a smaller CI list --- is less expensive than simply
doubling the sampling effort on the full Jastrow-CASCI wavefunction.
However, both these methods still rely upon sampling with and attempting to
optimize many CI parameters that are ultimately not helping.
This apparent inefficiency in mind, we now turn to
a scheme inspired by sCI that builds up from a small list
instead of pruning down from a large one.

\subsection{An sCI-inspired approach}
\label{sec:TrueSCI}
In this final section, we look to build a more traditional sCI inspired method for identifying statistically resolvable CI parameters.
Gradually increasing a variational subspace in the standard sCI way not only offers better computational economy, but should be more practical than our first two methods when moving beyond our small test system--for example, in larger molecule where even a conservative sCI calculation has order millions of determinants.

For this approach, we alternate between stages of VMC optimization and sCI-style subspace expansion.
We begin with only the Aufbau determinant and optimize the Jastrow parameters using our standard optimizer.
In the first CI expansion stage, all possible single and double excitations relative to Aufbau, but (for fair comparisons) within the same active space of the CASCI described above, are generated and added to the CI expansion with coefficient values of zero.
Excitations beyond this active space are certainly important for high accuracy, but here we are looking to make direct comparisons to our previous data and limit our search space to that of the CASCI wavefunction.
We then generate 24 estimates of the first column of the linear method Hamiltonian matrix, each with 250,000 samples.
Each determinant is selected or rejected from the next optimization based on its estimated Hamiltonian matrix element with the entirety of the just-optimized wavefunction, which is just its element in the first column of the LM Hamiltonian matrix.
We decide to include or exclude that individual parameter based on the ratio of the matrix element's average value to its standard deviation over the 24 LM Hamiltonian matrix columns.
In addition to the singles and doubles expansion candidates, we also test
the determinants already in our wavefunction and remove any that do not
pass the same criteria.
Once we have our updated CI expansion, we optimize the trial wavefunction
again.
At each subsequent expansion iteration, the candidates are generated as all unique possible single and double excitations relative to all determinants currently in the CI space (again limiting to the CASCI subspace).
Any determinant that was already present and survives to participate
in the next iteration carries its optimized coefficient forward.

\begin{figure*}
    \centering
    \includegraphics[width=\textwidth]{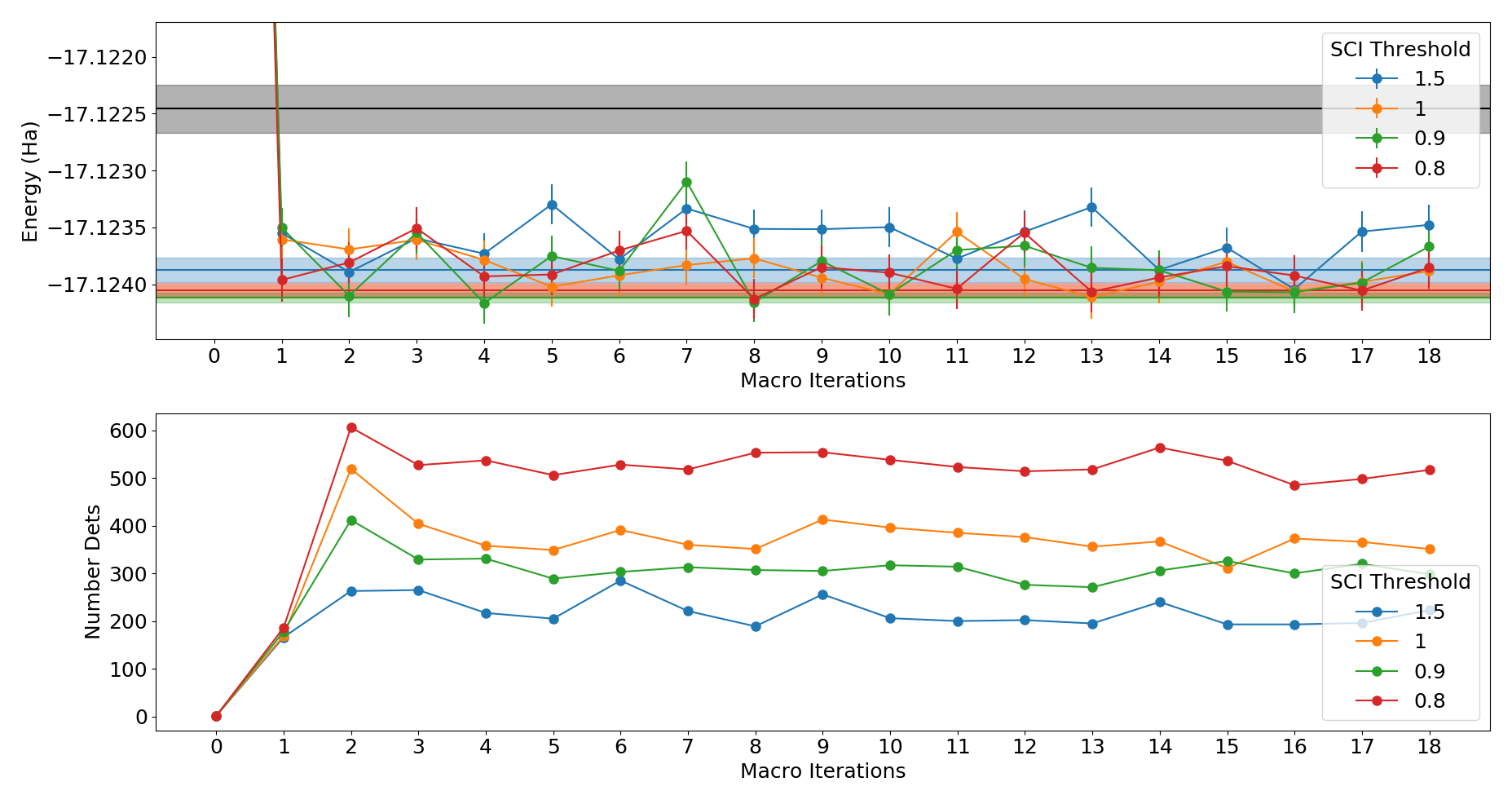}
    \caption{Selective CI optimizations with different selection thresholds (see text).  The top panel contains optimization energy progress for the different SCI calculations.  The first point of each colored line is the single Slater Jastrow wavefunction, which has energy about -17.104 Ha.  The shaded black line is the same average from Figure \ref{fig:10TrialData}. The shaded colored lines are the averages and standard deviations of the 5 lowest energy points for each respective color.  
    All optimizations are done using the standard optimizer sampling effort.
    The bottom panel shows the number of determinants at each macro iteration.}
    \label{fig:SCI_Data}
\end{figure*}

Figure \ref{fig:SCI_Data} shows our sCI optimization progress for a few different determinant selection thresholds.
An sCI threshold of 1 means the Hamiltonian matrix element's average must be greater than its standard deviation.
Likewise, a threshold of 1.5 means the matrix element's average value must be greater than 1.5 times its standard deviation.
No matter the threshold, by the completion of the first macro iteration (that is, after an optimization of the Jastrow parameters with only the Hartree Fock determinant, one stage of selection, and then optimizing the expanded wavefunction), all optimizations have lower variational energies than the average value from section \textbf{a)} of Figure \ref{fig:10TrialData}.
Notably, each of the VMC optimizations in this approach are done with our standard optimization with 1x sampling, meaning that two relatively cheap (due to low determinant number) optimizations can result in a lower variational energy than the 1x sampling optimizer applied to the full Jastrow-CASCI wavefunction.
Indeed, the energy at this point is already competitive with the 4x sampling approach to full Jastrow-CASCI. 
All thresholds add approximately the same number of determinants during this first macro iteration.

After the first iteration, for all thresholds, the energy bounces around, but never going above 1x full Jastrow-CASCI energy.
By about the third macro iteration, the number of determinants has approximately stabilized for all thresholds.
The number of determinants included follows the expected trend: when being very selective in what is included into the next optimization (by requiring a large ratio of Hamiltonian matrix element value to standard deviation, ie 1.5), the fewest determinants are added.
When being much more aggressive, with a threshold of 0.8, the most determinants are included.
The trend holds for the 0.9 and 1.0 thresholds relative to 1.5 and 0.8, although their ordering is flipped--the 0.9 threshold typically has fewer determinants than 1.0.
Likely, this is due to the stochastic nature of the selection scheme, and the fact that each wavefunction selection stage is dependent on the wavefunction previous, meaning any differences can be carried forward.

To further improve the efficiency of this scheme, we look to save time by altering our optimization schedule.
Utilizing the complete standard optimizer on a single determinant to optimize just one and two body Jastrows is excessive sampling effort!
Further, we recognize that by the completion of the 3rd macro iteration, all of our thresholds have plateaued in terms of the total number of determinants.
With these pieces in mind, we take the following approach--for the first four optimizations, we decrease the sampling effort of by 16x, 8x, 4x, and 2x, respectively, while keeping the number of total optimization iterations constant.
While these optimizations are certainly not going to tightly converge, we are simply looking to get a good enough approximation to the optimal parameter values, which by construction are those which are likely easily statistically resolvable.
Ideally, we will quickly reach the plateau of total number of determinants with minimal sampling effort, at which point we can increase our sampling effort to the standard level and optimize fully.

\begin{figure*}
    \centering
    \includegraphics[width=\textwidth]{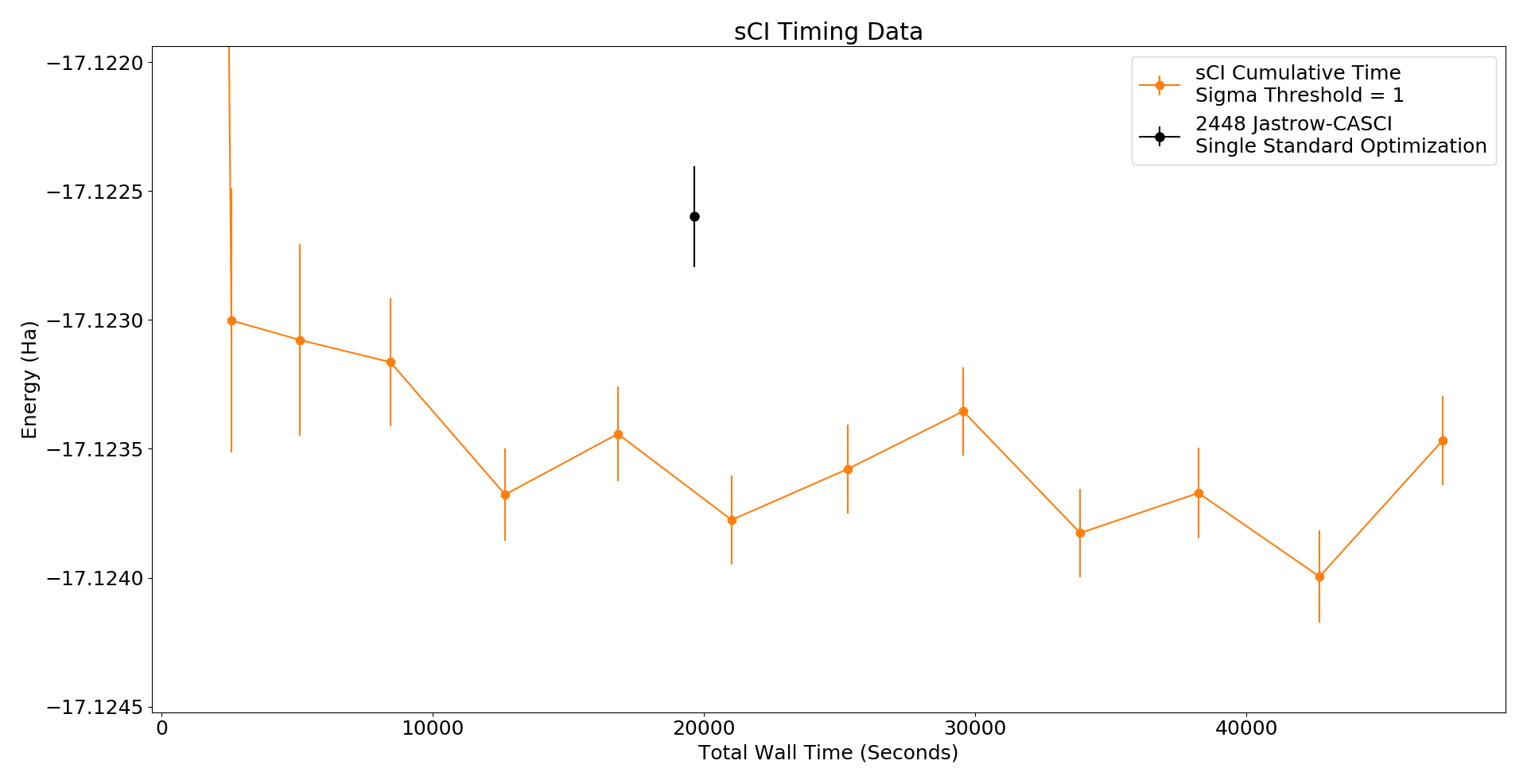}
    \caption{Timing for sCI, with sigma threshold of 1.  Each point is comprised of the total time for the SCI stage to determine that wavefunction and the following optimization.  The first four points utilize the standard optimization scheme, but with 1/16, 1/8, 1/4, and 1/2 the total samples at each individual iteration, respectively.  All other optimizations utilize the full sampling of the standard optimizer described in Table \ref{tab:optimizer_sampling_effort}.  A single optimization of the full CASCI wavefunction on identical hardware is plotted for comparison.}
    \label{fig:SCItiming}
\end{figure*}

The results of this method are shown in Figure \ref{fig:SCItiming}, along with a single complete Jastrow-CASCI optimization (section \textbf{a)} of Figure 1), with both calculations run on identical hardware.
For timing, each sCI point contains both the total optimization time along with the overhead time required to generate the 24 LM Hamiltonian estimates.
Even with the extra sampling during the SCI stage, we achieve an energy better than the  Jastrow-CASCI wavefunction with the standard optimization in less time.
Our energy is competitive with much more aggressive 4x and 8x sampling of the complete Jastrow-CASCI wavefunction.
Estimating the 4x and 8x sampling optimizations would have taken 4x and 8x the black point wall time places both of these optimizations well beyond the horizontal axis as plotted.
Further, this scheme is strictly faster than the median removal and both parameter statistical removal schemes, as each of those require first optimizing the complete Jastrow CASCI wavefunction before running more optimizations.
Clearly, using sCI ideas within VMC can offer advantages
for optimization speed and efficacy.

\section{Conclusions}
\label{sec:Conclusions}

We have presented an investigation into the interplay between
optimization efficiency and the ability to statistically resolve
CI parameters in VMC.
We find evidence suggesting that working to avoid statistically
unresolvable parameters can be more effective at improving
optimization results than simply increasing the sampling effort.
In particular, we tested two schemes for removing the least
statistically significant CI parameters from an optimized VMC wavefunction,
as well as a scheme that adds parameters based on the statistical
significance of their sCI-style matrix elements.
All of these schemes were able to produce better energies than
a traditional VMC optimization at similar or reduced cost.

Looking forward, we expect that a more carefully tuned combination
of these schemes will offer significant advantages in the
optimization of large CI expansions within VMC.
Beyond CI parameters, it would be interesting to study whether
orbital, Jastrow, and backflow parameters show similar
relationships between statistical uncertainty and optimization
efficacy.
It will also be important to establish clear and well-balanced
stopping criteria for VMC optimizations in which the active
variational parameters are determined on the fly.
Energy variance targets are one option, but better options
may exist.
Certainly, methods that can alleviate the computational
burden of large sample sizes and help determine the best
form of wavefunction to use at a given level of effort
will be welcome.

\section{Appendix}
\subsection{Water Geometry}
\label{sec:WaterGeom}
\begin{table}[h]
    \centering
    \begin{tabular}{lrrr}
\hline
\hline
\multicolumn{4}{c}{All units are Angstrom} \\
\hline
  O &  0.000000 &   0.000000 &  0.117790 \\
  H &  0.000000 &   0.755453 & -0.471161 \\
  H &  0.000000 & -1.1331795 &  -0.7067415 \\
\hline
    \end{tabular}
    \caption{Water Geometry}
    \label{tab:water_geom}
\end{table}

\subsection{Figure 1 Optimization Details}
\label{sec:Fig1OptimizationDetailsTable}
\begin{table*}[h]
    \centering
    \begin{tabular}{|c|l|c|r|r|r|}
        \hline
        Section Label in Figure \ref{fig:10TrialData} & Optimizer & Macro iterations & Optimization Method  & Micro iterations & Samples (per micro) \\
        \hline
        \hline
        \multirow{3}{*}{\textbf{a)}} &
        \multirow{2}{*}{Hybrid Method} & \multirow{2}{*}{16} & Accelerated Descent & 100 & 30,000 \\
        &&&Block Linear Method & 3 & 500,000 \\      
        &Descent Finalization & 1 & Accelerated Descent & 1000 & 100,000 \\
        \hline
        \multirow{3}{*}{\textbf{b)}} &
        \multirow{2}{*}{Hybrid Method} & \multirow{2}{*}{8} & Accelerated Descent & 100 & 60,000 \\
        &&&Block Linear Method & 3 & 1,000,000 \\      
        &Descent Finalization & 1 & Accelerated Descent & 500 & 200,000 \\
        \hline
        \multirow{3}{*}{\textbf{c)}} &
        \multirow{2}{*}{Hybrid Method} & \multirow{2}{*}{8} & Accelerated Descent & 100 & 30,000 \\
        &&&Block Linear Method & 3 & 500,000 \\      
        &Descent Finalization & 1 & Accelerated Descent & 2000 & 100,000 \\
        \hline
        \multirow{3}{*}{\textbf{d)}} &
        \multirow{2}{*}{Hybrid Method} & \multirow{2}{*}{16} & Accelerated Descent & 100 & 30,000 \\
        &&&Block Linear Method & 3 & 500,000 \\      
        &Descent Finalization & 1 & Accelerated Descent & 2000 & 100,000 \\ 
        \hline
        \multirow{3}{*}{\textbf{e)}} &
        \multirow{2}{*}{Hybrid Method} & \multirow{2}{*}{16} & Accelerated Descent & 100 & 60,000 \\
        &&&Block Linear Method & 3 & 1,000,000 \\      
        &Descent Finalization & 1 & Accelerated Descent & 1000 & 200,000 \\         
        \hline
        \multirow{3}{*}{\textbf{f)}} &
        \multirow{2}{*}{Hybrid Method} & \multirow{2}{*}{32} & Accelerated Descent & 100 & 30,000 \\
        &&&Block Linear Method & 3 & 500,000 \\      
        &Descent Finalization & 1 & Accelerated Descent & 2000 & 100,000 \\                 
        \hline
        \multirow{3}{*}{\textbf{g)}} &
        \multirow{2}{*}{Hybrid Method} & \multirow{2}{*}{16} & Accelerated Descent & 100 & 30,000 \\
        &&&Block Linear Method & 3 & 500,000 \\      
        &Descent Finalization & 1 & Accelerated Descent & 3621 & 100,000 \\               
        \hline
        \multirow{3}{*}{\textbf{h)}} &
        \multirow{2}{*}{Hybrid Method} & \multirow{2}{*}{64} & Accelerated Descent & 100 & 30,000 \\
        &&&Block Linear Method & 3 & 500,000 \\      
        &Descent Finalization & 1 & Accelerated Descent & 2000 & 100,000 \\   
        \hline
        \multirow{3}{*}{\textbf{i)}} &
        \multirow{2}{*}{Hybrid Method} & \multirow{2}{*}{16} & Accelerated Descent & 100 & 120,000 \\
        &&&Block Linear Method & 3 & 2,000,000 \\      
        &Descent Finalization & 1 & Accelerated Descent & 1000 & 400,000 \\         
        \hline
        \multirow{3}{*}{\textbf{j)}} &
        \multirow{2}{*}{Hybrid Method} & \multirow{2}{*}{16} & Accelerated Descent & 100 & 240,000 \\
        &&&Block Linear Method & 3 & 4,000,000 \\      
        &Descent Finalization & 1 & Accelerated Descent & 1000 & 800,000 \\         
        \hline
    \end{tabular}
    \caption{Details of each of the optimizers used in Figure \ref{fig:10TrialData}.}
    \label{tab:all_optimizers_fig_1}
\end{table*}


%
%

%


\bibliography{SCIVMC}

\end{document}